\documentclass[aps,twocolumn,showpacs,preprintnumbers,amsmath,amssymb,floats,citeautoscript,nobalancelastpage]{revtex4-1}
\usepackage{graphicx}
\usepackage{dcolumn}
\usepackage{bm}
\usepackage{txfonts}
\usepackage{multirow}
\usepackage{color}

\begin{document}
 
\title{Thermoelectric transport in the layered Ca$_3$Co$_{4-x}$Rh$_{x}$O$_9$ single crystals}

\author{Yusuke~Ikeda}
\author{Kengo~Saito}
\author{Ryuji~Okazaki$^{\ast}$}

\affiliation{Department of Physics, Faculty of Science and Technology, Tokyo University of Science, Noda 278-8510, Japan}

\begin{abstract}
We have examined an isovalent Rh substitution effect on the transport properties of 
the thermoelectric oxide Ca$_3$Co$_{4}$O$_9$
using single-crystalline form.
With increasing Rh content $x$, 
both the electrical resistivity and the Seebeck coefficient 
change systematically  up to $x=0.6$ for Ca$_3$Co$_{4-x}$Rh$_{x}$O$_9$ samples.
In the Fermi-liquid regime where the resistivity behaves as $\rho=\rho_0+AT^2$ around 120~K, 
the $A$ value decreases with increasing Rh content, indicating that 
the correlation effect is weakened by Rh $4d$ electrons with extended orbitals.
We find that,
in contrast to such a weak correlation effect observed in the resistivity of Rh-substituted samples,
the low-temperature Seebeck coefficient is increased with increasing Rh content,
which is explained with a possible enhancement of a pseudogap associated with the short-range order of spin density wave.
In  high-temperature range above room temperature,
we show that 
the resistivity is largely suppressed by Rh substitution
while the Seebeck coefficient becomes almost temperature-independent,
leading to a significant improvement of the power factor in Rh-substituted samples.
This result is also discussed in terms of the differences in the orbital size and the associated spin state 
between Co $3d$ and Rh $4d$ electrons.

\end{abstract}

\maketitle

\section{introduction}

Thermoelectric material, which can directly interconvert thermal and electrical energy
through the Seebeck and Peltier effects,
has attracted increasing attention
as a solution to realize a sustainable and emission-free society
without environmental pollution \cite{Snyder2008,Bell2008}.
It offers simple and compact solid-state energy-conversion devices 
that are adaptive in a variety of situations 
to utilize various kinds of waste heat
such as car engine and garbage furnace.
In this context, oxide thermoelectrics is
potentially promising 
because it is fairly stable at high temperatures in air,
and since the discovery of the good thermoelectric properties of
the layered cobalt oxide Na$_x$CoO$_2$ \cite{Terasaki1997},
oxide thermoelectrics including  transition-metal elements
has been intensively studied in the past decade \cite{Maignan2002,Koumoto2010,He2011}.

Among the oxide thermoelectrics,
the layered Ca$_3$Co$_{4}$O$_9$ is a peculiar example
because of 
its highly unusual transport behaviors 
intimately related to the complicating magnetic properties.
Ca$_3$Co$_{4}$O$_9$
consists of alternate stacking of 
two types of layered subsystems with different $b$-axis parameter (Fig. 1) \cite{Masset2000}:
The CdI$_2$-type CoO$_2$ layer responsible for the charge transport
and
the rocksalt-type Ca$_2$CoO$_3$ layer
which behaves as the
charge reservoir to supply holes into the CoO$_2$ layer \cite{Yang2008,Tanabe2016,Mizokawa2005,Klie2012}.
The anomalous charge transport of this compound is clearly seen in 
the non-monotonic temperature variations of 
the electrical resistivity and the Seebeck coefficient \cite{Masset2000,Miyazaki2000,Shikano2003,Limelette2005,Hejt2015},
but an underlying origin for these behaviors remains unclear.
For instance, below $T\sim60$~K, the resistivity increases with decreasing temperature
like an insulator while the Seebeck coefficient shows a metallic behavior,
which has been discussed in terms of the variable range hopping \cite{Bhaskar2014},
quantum criticality \cite{Limelette2010},
and
pseudogap opening \cite{Hsieh2014} 
associated with a spin-density-wave (SDW) formation \cite{Sugiyama2002}.
At higher temperature, 
the resistivity shows a Fermi-liquid behavior varying as $\rho(T)=\rho_0+AT^2$ 
up to the coherence temperature $T^*\sim140$~K, above which 
an incoherent bad-metal state 
is realized.
The Seebeck coefficient reaches a large value of 130~$\mu$V/K at around 150~K, and
becomes almost temperature independent up to 400~K,
as described within
the modified Heikes formula in which 
the spin and orbital degrees of freedom of correlated Co $3d$ electrons
are involved \cite{Koshibae2000,Koshibae2001}.
On the other hand, the Seebeck coefficient again increases 
with increasing temperature above 400~K,
suggested to be associated with a spin-state crossover \cite{Shikano2003},
but the high-temperature spin-state nature of Ca$_3$Co$_{4}$O$_9$ remains puzzling \cite{Sugiyama2003,Wakisaka2008}.
Note that the similar issue regarding an excited spin state is greatly debated in the perovskite LaCoO$_3$ 
\cite{Raccah1967,Asai1989,Korotin1996,Haverkort2006}.

\begin{figure}[b]
\includegraphics[width=1\linewidth]{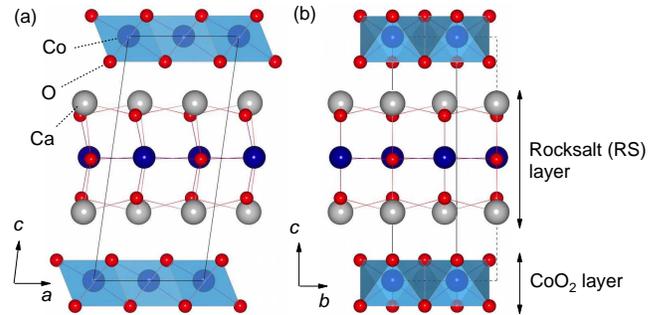}
\caption{(Color online).
Crystal structure of Ca$_3$Co$_{4}$O$_9$ viewed from (a) $b$-axis and (b) $a$-axis directions.
The solid and dashed lines in (b) respectively represent the unit cells for CoO$_2$ and rocksalt layers
with different $b$ axis parameters.
}
\end{figure}

To clarify a precise role of correlated Co $3d$ electrons for 
the intriguing electronic properties in this compound,
it is quite valuable to substitute the Co ions by
isovalent transition-metal elements  
with different orbital size and correlation effect \cite{Klein2006,Shibasaki2011,Okazaki2011}.
The physical properties of 
 polycrystalline samples of isovalent Rh-substituted Ca$_3$Co$_{4}$O$_9$ 
have been recently examined for this purpose \cite{Huang2013}, however,
the polycrystalline samples
have an unavoidable extrinsic effect on the charge transport such as the boundary scattering.
Also note that the anisotropy of the resistivity in the parent compound is significantly large  \cite{Masset2000}.

The aim of this paper is to experimentally investigate how an extended Rh 4$d$ orbital affects 
the transport properties of Ca$_3$Co$_{4}$O$_9$
using single-crystalline samples.
As claimed in the previous polycrystalline study \cite{Huang2013},
in the low-temperature Fermi-liquid regime, 
the $A$ value in the resistivity is reduced by Rh substitution, indicating that 
an electron-electron correlation effect is weakened by Rh $4d$ electrons with broad orbitals.
We find that,
in contrast to such a weak correlation effect,
the low-temperature Seebeck coefficient is unexpectedly enhanced
in the Rh-substituted single crystals,
which may attribute to 
an enlargement of the pseudogap in the Rh-substituted samples 
associated with the short-range order of SDW.
In the incoherent transport regime above room temperature,
the resistivity is reduced by Rh substitution,
which also stems from the extended  $t_{2g}$ orbitals of Rh $4d$ electrons with light effective mass.
The Seebeck coefficient of the Rh-substituted single crystals is found to become almost temperature-independent
at high temperatures, 
qualitatively understood from the spin state of Rh ions, which favors the low-spin state
owing to a large crystal field splitting for the $4d$ electrons.

\section{experiments}

The experiments were performed using 
Ca$_3$Co$_{4-x}$Rh$_{x}$O$_9$ single crystals ($x=0,0.2,0.4,0.6$) with typical sample dimensions of $1\times1\times0.02$\,mm$^3$ 
grown by a flux method \cite{Mikami2006}. 
Precursor powders of CaCO$_3$ (99.9\%), Co$_3$O$_4$ (99.9\%), and Rh$_2$O$_3$ (99.9\%) 
were mixed in a stoichiometric ratio and
calcined two times in air at 1173~K for 24~h with intermediate grindings.
Then KCl  (99.999\%) and K$_2$CO$_3$  (99.999\%) powders mixed with a molar ratio of $4:1$ was added 
with the calcined powders of Ca$_3$Co$_{4-x}$Rh$_{x}$O$_9$ as a flux. 
The concentration of Ca$_3$Co$_{4-x}$Rh$_{x}$O$_9$ was set to be 1.5\% in molar ratio. 
The mixture was put in an alumina crucible and 
heated up to 1123~K in an electrical furnace with a heating rate of 200~K/h.
After keeping the maximum temperature for 1~h, 
the furnace was cooled down slowly with a rate of 1~K/h, 
and at 1023~K, the electrical power of the furnace was switched off.
As-grown samples were rinsed in distilled water to remove the flux and
platelet crystals with a shiny surface were obtained.

The x-ray diffraction measurement was performed with Cu $K\alpha$ radiation
in a $\theta$-$2\theta$ scan mode.
The scattering vector was normal to the crystal surface. 
The in-plane resistivity was measured using a standard four-probe method.
Excitation current of $I=20$~$\mu$A was provided by a Keithley 6221 current source and the sample voltage
was measured with a synchronized Keithley 2182A nanovoltmeter. 
These two instruments were operated in a built-in Delta mode to cancel unwanted thermoelectric voltage.
The in-plane Seebeck coefficient was measured using a steady-state technique
with a typical temperature gradient of 0.5 K/mm made by a resistive heater.
The thermoelectric voltage of the sample was measured with Keithley 2182A nanovoltmeter.
The temperature gradient was measured with
a differential thermocouple made of copper and constantan in a liquid He cryostat from 4~K to 300~K and
with two platinum temperature sensors in a high-temperature equipment from 300~K to 650~K.
In both measuring apparatuses,
the thermoelectric voltage from the wire leads was carefully subtracted.

\section{results}

\begin{figure}[t]
\includegraphics[width=1\linewidth]{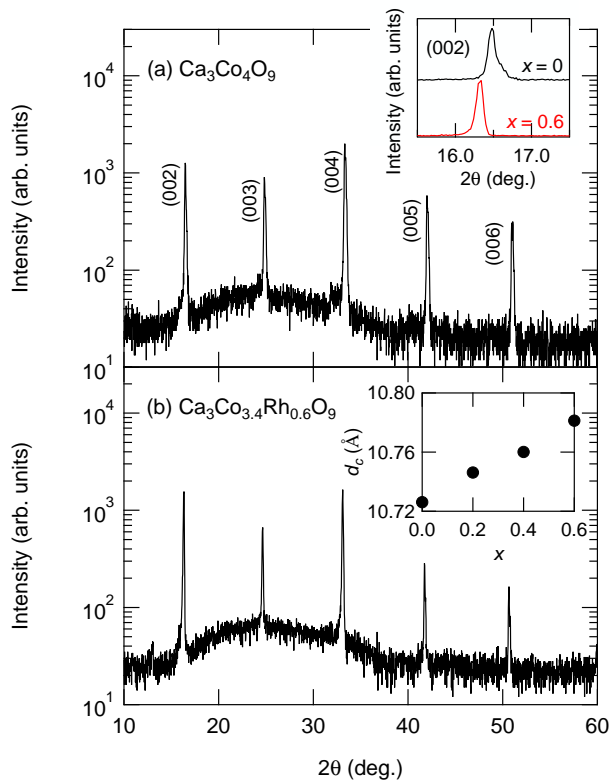}
\caption{(Color online).
X-ray diffraction patterns of the as-grown single crystals of 
(a) Ca$_3$Co$_{4}$O$_9$ and
(b) Ca$_3$Co$_{3.4}$Rh$_{0.6}$O$_9$.
Inset (a) compares these two spectra around the (002) reflection.
The interplanar distance $d_c$ of Ca$_3$Co$_{4-x}$Rh$_{x}$O$_9$ 
determined from $(00l)$ peaks is shown in the inset (b)
as a function of Rh content $x$.
}
\end{figure}

Figures 2(a) and 2(b) represent typical 
x-ray diffraction patterns of the as-grown single crystals of 
Ca$_3$Co$_{4}$O$_9$ ($x=0$) and Ca$_3$Co$_{3.4}$Rh$_{0.6}$O$_9$ ($x=0.6$), respectively,
measured with the scattering vector normal to the surface of the plate-like samples. 
All the peaks are indexed as $(00l)$ reflections of the parent compound,
meaning that the measured crystal surface is the $ab$ plane.
In Bi-substituted Ca$_3$Co$_{4}$O$_9$, it is reported that an impurity phase is grown even in single crystals \cite{Mikami2006}.
The present crystals do not show such impurity phase even for $x=0.6$ sample.
The inset of Fig. 2(a) compares the diffraction patterns of 
$x=0$ and $x=0.6$ samples around the (002) reflection.
This (002) peak is shifted low angle with increasing Rh content $x$
owing to the larger ionic radius of Rh than that of Co.
Since the crystal structure of Ca$_3$Co$_{4}$O$_9$ is 
monoclinic ($C2/m$), 
it is impossible to determine the $c$-axis parameter 
from only  $(00l)$ peaks. 
Instead, we have obtained the interplanar distance
$d_c$ from $2d_c\sin\theta=n\lambda$,
where $n$ is integer and $\lambda$ is the wavelength of incident x-ray,
by using all the $(00l)$ reflections.
As shown in the inset of Fig. 2(b),
$d_c$ linearly increases as a function of $x$,
indicating successful Rh substitutions.
At present,
it is unclear 
whether Co site of CoO$_2$ or the rocksalt layer is substituted by Rh ions.
While
the precise position of Rh ions should be studied in future
spectroscopic experiments,
a considerable and systematic change in the following transport properties
implies that Rh ions may substantially enter the conductive CoO$_2$ layer.

\begin{figure}[t]
\includegraphics[width=1\linewidth]{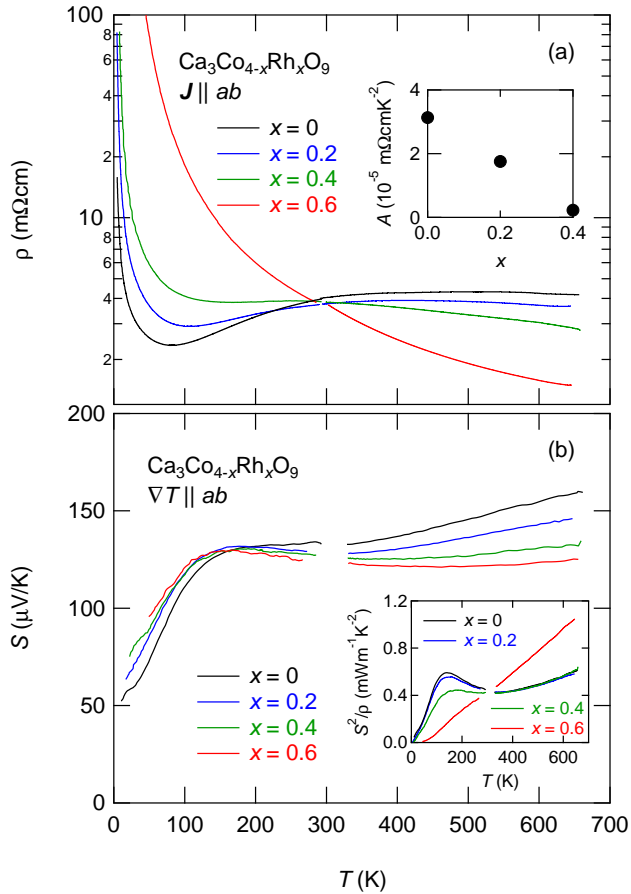}
\caption{(Color online).
Temperature variations of (a) the resistivity and (b) the Seebeck coefficient of
Ca$_3$Co$_{4-x}$Rh$_{x}$O$_9$ single crystals ($x=0, 0.2, 0.4, 0.6$).
The Fermi-liquid transport coefficient $A$ is shown in the inset of (a) as
a function of Rh content $x$.
Inset of (b) shows the temperature variations of the power factor $S^2/\rho$.
}
\end{figure}

Figure 3(a) shows the temperature variations of 
the electrical resistivity measured along the in-plane 
direction.
Around room temperature, the in-plane resistivity 
is about 4~m$\Omega$cm for all the samples,
which is the same as the single-crystalline result of
the parent compound \cite{Limelette2005}.
Below room temperature
it is largely increased with increasing Rh content $x$.
For $x=0,0.2,0.4$ samples, the resistivity decreases 
with decreasing temperature down to around 100~K 
and then 
shows an upturn towards lower temperatures.
In the metal-like regions around 120~K, 
the resistivity exhibits
a Fermi-liquid behavior varying as $\rho(T)=\rho_0+AT^2$,
and the $A$ value decreases with Rh content $x$ 
as displayed in the inset of Fig. 3(a).
This behavior is also reported in the previous study on the 
polycrystalline samples \cite{Huang2013},
although the absolute value is quite different because of the 
single-crystalline nature in the present study.
Since $A$ can be scaled to $m^{*2}$, where $m^*$ is the effective mass \cite{Kadowaki1986},
this result indicates that the effective mass becomes small with Rh substitutions,
as is naturally understood from a weak correlation effect of Rh $4d$ electrons compared 
with Co $3d$ electrons.
For $x=0.6$ sample, such a Fermi-liquid behavior
is no longer observed.

In contrast to the low-temperature results, the resistivity decreases 
with increasing Rh content
above room temperature.
The resistivity is less than 2~m$\Omega$cm for  $x=0.6$ sample
at 600~K, which is beneficial for oxide thermoelectrics at high temperatures
as discussed later.
In the parent compound, the resistivity shows a broad maximum around 500~K
and then decreases with heating.
With increasing Rh content, the broad peak shifts to the lower temperature,
and disappears for  $x=0.6$ sample,
in which the resistivity shows an insulating behavior in the whole temperature
range we measured.
We infer that 
such an insulating behavior is ascribed not to
the carrier concentration, which is essentially constant above 
room temperature,
but to the mobility,
as will also be discussed in the results of the Seebeck coefficient.
Such a nonmetallic resistivity due to the mobility has also
been observed in several oxide materials \cite{Shibasaki2006,Takahashi2012}.

Figure 3(b) displays 
the temperature dependence of the in-plane  Seebeck coefficient.
Below room temperature,
the Seebeck coefficient slightly but systematically changes with Rh
content;
In particular, below 100~K where the Seebeck coefficient shows
a metallic behavior, it is considerably enhanced by Rh substitution.
The present results are distinct from the 
previous report using polycrystalline samples which shows 
that Rh substitution does not affect the Seebeck coefficient \cite{Huang2013},
probably attributed to an extrinsic effect of averaging large transport anisotropy 
in the polycrystalline samples.
Above room temperature,
the enhancement of the Seebeck coefficient seen in the parent compound is 
significantly suppressed with Rh substitution.
For $x=0.6$ sample, the Seebeck coefficient becomes almost temperature independent 
up to 650~K,
where the Heikes formula with a constant carrier density looks applicable.

\section{discussion}

\subsection{Low-temperature transport below 300~K}

Let us discuss an origin of the transport behaviors 
in this system.
We first focus on the low-temperature upturn of the resistivity 
coexisting with a metallic Seebeck coefficient below 100~K.
Although it has been often discussed in 
terms of variable range hopping due to carrier localization \cite{Bhaskar2014},
it is claimed that the temperature dependence of the resistivity
is much milder than that of the variable-range-hopping transport $\exp(T^{-\alpha})$
($0<\alpha<1$) \cite{Hsieh2014}
and that the fitting range is narrow.
A scenario of quantum criticality is quite intriguing \cite{Limelette2005},
but it is difficult to explain the Seebeck coefficient of the Rh-substituted samples in following reason;
As mentioned before, the $A$ value in the resistivity, a measure of strength of the correlation effect,
decreases with increasing Rh content, as is understood from a 
weak correlation of Rh $4d$ electrons.
This indicates that the Rh-substituted samples locate far from the quantum critical point.
Since the $A$ value is intimately related to the 
electronic specific heat and Seebeck coefficient through the effective mass \cite{Behnia2004},
the Rh-substituted samples with lower effective mass should have a smaller Seebeck
coefficient than that of the parent Ca$_3$Co$_{4}$O$_9$ in the case of the quantum critical scenario.
This is, however, opposite to the present results in which the Seebeck coefficient is increased by Rh substitution 
at low temperatures.

As an origin of the low-temperature transport behaviors,
we consider a formation of the pseudogap associated with the short-range order of 
SDW, which is suggested in 
Bi-substituted Ca$_3$Co$_{4}$O$_9$ single crystals \cite{Hsieh2014}.
At low temperatures,
the Seebeck coefficient in a two-dimensional metal is expressed as
\begin{equation}
S=\frac{\pi k_B^2}{2q\hbar^2 c_0}\frac{m^*}{n}T,
\end{equation}
where $q$ is the charge of carrier, $c_0$ the $c$-axis length,  
$n$ the carrier concentration \cite{Kresin1990}.
Now the conduction carrier is hole, 
hence $q=+e$.
Using Eq. (1), $n/m^*$ can be evaluated from $T/S$ as
\begin{equation}
\frac{T}{S}=
\frac{2e\hbar^2 c_0}{\pi k_B^2}\frac{n}{m^*}.
\end{equation}
Figure 4 depicts the temperature dependence of 
$T/S$ 
at low temperatures.
Hsieh $et$ $al$ have shown that 
the decrease of $T/S$ with lowering temperature 
originates from the  decrease of  the carrier concentration $n$ 
according to Eq. (2) \cite{Hsieh2014},
which is also supported from the results of the Hall coefficient measurements \cite{Eng2006},
and proposed that this phenomenon is ascribed to 
an opening of the pseudogap
caused by the short-range order of SDW below 100~K \cite{Sugiyama2002}.
We also note that such a reduction of $T/S$ below a pseudogap temperature is observed in 
other layered cobalt oxide \cite{Itou2000} and   
the Kondo semiconductor CeNiSn \cite{Nakamoto1995}.

\begin{figure}[t]
\includegraphics[width=1\linewidth]{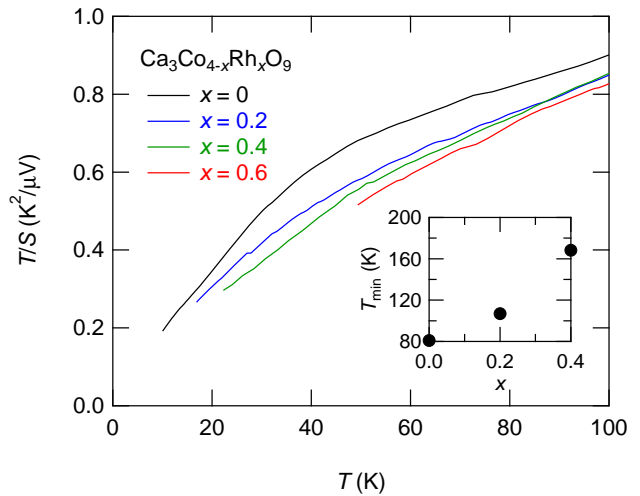}
\caption{(Color online).
Temperature variations of $T/S$ of
Ca$_3$Co$_{4-x}$Rh$_{x}$O$_9$ single crystals ($x=0, 0.2, 0.4, 0.6$)
at low temperatures.
The inset depicts the Rh content $x$ dependence of $T_{min}$
at which the resistivity takes the minimum value
in the low-temperature range.
}
\end{figure}

We then discuss how the Rh substitution affects the 
pseudogap behavior.
In the present study, $T/S$ systematically decreases with Rh substitution
below 100~K.
Within the pseudogap scenario,
the present result shows that $n$ decreases with increasing Rh content,
indicating that the pseudogap associated with SDW becomes 
large by Rh substitution.
Now we stress that this is consistent with our resistivity results.
In this system, the resistivity minimum around 70~K is
associated with an onset of SDW gap formation \cite{Kobayashi2006,Wang2010}.
Therefore, according to above scenario,
the onset temperature should increase with increasing Rh content.
The inset of Fig. 4 displays the Rh content dependence of  
$T_{min}$
at which the resistivity takes the minimum value
in the low-temperature range.
Indeed, the onset temperature $T_{min}$ increases with increasing Rh content,
supporting the pseudogap scenario.
On the other hand,
it is an open question at present why the pseudogap is stabilized by Rh substitution.
An underlying mechanism should be examined by future  microscopic measurements.

\subsection{High-temperature transport above 300~K}

Next we remark the high-temperature transport in this system.
In the parent compound, the Seebeck coefficient around room temperature is 
well explained by the modified Heikes formula \cite{Koshibae2000,Koshibae2001},
which considers  
a hole hopping between Co$^{3+}$ and Co$^{4+}$ sites
both of which take low-spin states \cite{Mizokawa2005,Klie2012}.
In contrast, an origin of the enhancement of the Seebeck coefficient above 400 K is not clear.
It is theoretically suggested that a mixture of low-spin and high-spin states may enhance the Seebeck 
coefficient \cite{Koshibae2000}. 
Thus, a thermally-induced spin-state crossover from low-spin into higher-spin state is suggested for the enhancement of Seebeck coefficient \cite{Shikano2003}, 
but the nature of spin states at high temperatures is still puzzling \cite{Sugiyama2003,Wakisaka2008}.

In the present study, the Rh substitution  suppresses such an enhancement,
and the Seebeck coefficient becomes almost temperature independent for $x=0.6$ sample.
This behavior may be naively understood within a suppression of 
the spin-state crossover by Rh substitution.
In contrast to Co ions, 
the spin state of which often dramatically varies with temperature,
Rh ions usually show the stable low-spin state
owing to the large crystal field splitting between
$t_{2g}$ and $e_g$ manifolds \cite{Furuta2014}.
Thus, at high temperatures, 
the number of the transition-metal ions with higher-spin states,
which can enhance the Seebeck coefficient,
would be decreased in average by increasing Rh content.
This leads to a possible suppression of the high-temperature Seebeck coefficient in Rh-substituted crystals.

Let us discuss an origin of the reduction of the high-temperature resistivity in the Rh-substituted samples.
In the high-temperature range, the Seebeck coefficient is described by the Heikes formula,
indicating that the carrier concentration is temperature-independent.
We also note that
it shows almost the same value of 130~$\mu$V/K from 200 to 300~K
for all the samples.
This implies that the carrier concentration does not change 
with Rh substitution, which is also supported by the isovalent substitution.
Therefore the decrease of the resistivity in Rh-substituted samples originates 
from an increase of hole mobility by Rh substitution.
Here, the hole hopping occurs in the $t_{2g}$ manifolds of Co/Rh ions,
and 
the effective mass of carriers becomes small with increasing Rh content
because of 
the weak correlation and the broad orbital of the Rh $4d$ electrons.
This may lead to a higher mobility in the Rh-substituted crystals.

We also mention a bad-metal state above $T^*\sim140$~K realized in the parent compound.
In a bad-metal state, 
a hallmark of strongly correlated electron systems \cite{Takenaka2002},
the resistivity is increased with increasing temperature beyond the Mott limit.
In the present study, as seen in the Fermi-liquid parameter $A$ in the resistivity,
the correlation effect is weakened in Rh-substituted samples and 
consequently the bad-metal state disappears in $x=0.6$ sample.
Since the resistivity increases in the bad-metal state,
the disappearance of bad-metal state,
along with the increase of hole mobility discussed above, may lead to a reduction of the resistivity in $x=0.6$ sample.

Finally let us  comment on the thermoelectric energy-conversion efficiency of the present crystals.
The inset of Fig. 3(b) shows the temperature variations of the power factor
$S^2/\rho$ of the present crystals.
For $x=0.6$ sample, 
the power factor reaches a large value of $1\times10^{-3}$~W/mK$^2$ at 600~K
thanks to the low resistivity.
It is interesting to enlarge the power factor by decreasing effective mass
in contrast to the usual approach to the high thermoelectric performance \cite{Pei2012}. 
Moreover, previous report using the polycrystalline samples shows that
the Rh substitutions strongly suppress the thermal conductivity in this system \cite{Huang2013}.
Thus the thermal conductivity of the  single-crystalline samples should be investigated
as a future issue.
An importance of the thermoelectric study using single crystals is also highlighted by 
the discovery of the high thermoelectric efficiency in SnSe single crystals \cite{Zhao2014}.

\section{summary}

We study the electrical and thermoelectric transport properties 
of the layered  Ca$_3$Co$_{4-x}$Rh$_{x}$O$_9$ single crystals
in a broad temperature range from 10~K up to 650~K.
In the metal-like region around 120~K, 
the Fermi-liquid transport coefficient $A$  decreases with increasing Rh content $x$,
indicating smaller effective mass in the Rh-substituted samples
because of the weak correlation effect of Rh $4d$ electrons compared 
with Co $3d$ electrons.
At lower temperatures,
in contrast to the weak correlation effect found in the Rh-substituted samples,
we observe that
the Seebeck coefficient increases with increasing Rh substitution,
which is also different from the results of the previous polycrystalline study.
We suggest that this behavior is attributed to an enlargement of a pseudogap 
in the Rh-substituted crystals, which is
accompanied 
with the short-range order of the spin density wave below 100~K,
as is supported from the increase of the spin-density-wave onset 
temperature by Rh substitution.

At higher temperatures, the resistivity is decreased with increasing 
Rh content owing to a small effective mass of Rh-substituted samples.
Such a low resistivity in Rh-substituted crystal indeed enhances the power factor
at high temperatures.
The Seebeck coefficient of Ca$_3$Co$_{3.4}$Rh$_{0.6}$O$_9$  
becomes almost temperature independent up to 650~K,
possibly originating from a suppression of a spin-state crossover
by the substitution of Rh ions which favor 
the low-spin state.
The microscopic measurements to elucidate the spin state
in the present system are highly required as a future issue.

\section*{acknowledgements}

This work was supported by 
a Grant-in-Aid for 
Challenging Exploratory Research (No. 26610099)
from JSPS.


\begin{thebibliography}{99}

\item[$^\ast$] Email: okazaki@rs.tus.ac.jp


\bibitem{Snyder2008}
G. J. Snyder and E. S. Toberer, 
Nat. Mater. {\bf 7}, 105 (2008).

\bibitem{Bell2008}
L. E. Bell,
Science {\bf 321}, 1457 (2008).

\bibitem{Terasaki1997} 
I. Terasaki, Y. Sasago, and K. Uchinokura,
Phys. Rev. B {\bf 56}, R12685 (1997).

\bibitem{Maignan2002}
A. Maignan, L. B. Wang, S. H\'ebert, D. Pelloquin, and B. Raveau, 
Chem. Mater. {\bf 14}, 1231 (2002).


\bibitem{Koumoto2010}
K. Koumoto, Y. Wang, R. Zhang, A. Kosuga, and R. Funahashi,
Ann. Rev. Mater. Res. {\bf 40}, 363 (2010).

\bibitem{He2011}
J. He, Y. Liu, and R. Funahashi, 
J. Mater. Res. {\bf 26}, 1762 (2011). 



\bibitem{Masset2000}
A. C. Masset, C. Michel, A. Maignan, M. Hervieu, O. Toulemonde, F. Studer, B. Raveau, and J. Hejtmanek, 
Phys. Rev. B {\bf 62}, 166 (2000).


\bibitem{Mizokawa2005}
T. Mizokawa, L. H. Tjeng, H.-J. Lin, C. T. Chen, R. Kitawaki, I. Terasaki, S. Lambert, and C. Michel,
Phys. Rev. B {\bf 71}, 193107 (2005).

\bibitem{Yang2008}
G. Yang, Q. Ramasse, and R. F. Klie,
Phys. Rev. B {\bf 78}, 153109 (2008).

\bibitem{Klie2012}
R. F. Klie, Q. Qiao, T. Paulauskas, A. Gulec, A. Rebola, S. \"O\ifmmode \breve{g}\else \u{g}\fi{}\"ut, M. P. Prange, J. C. Idrobo, S. T. Pantelides, S. Kolesnik, B. Dabrowski, M. Ozdemir, C. Boyraz, D. Mazumdar, and A. Gupta,
Phys. Rev. Lett. {\bf 108}, 196601 (2012).



\bibitem{Tanabe2016}
K. Tanabe, R. Okazaki, H. Taniguchi, and I. Terasaki,
J. Phys. Condens. Matter {\bf 28}, 085601 (2016).

\bibitem{Shikano2003}
M. Shikano and R. Funahashi,
Appl. Phys. Lett. {\bf 82}, 1851 (2003).

\bibitem{Miyazaki2000}
Y. Miyazaki, K. Kudo, M. Akoshima, Y. Ono, Y. Koike, and T. Kajitani,
Jpn. J. Appl. Phys. {\bf 39},  L531 (2000).

\bibitem{Limelette2005}
P. Limelette, V. Hardy, P. Auban-Senzier, D. J\'erome, D. Flahaut, S. H\'ebert, R. Fr\'esard, C. Simon, J. Noudem, and A. Maignan, 
Phys. Rev. B {\bf 71}, 233108 (2005).

\bibitem{Hejt2015}
J. Hejtm\'anek, Z. Jir\'ak, and J.  \ifmmode \check{S}\else \v{S}\fi{}ebek,
Phys. Rev. B {\bf 92}, 125106 (2015).


\bibitem{Bhaskar2014}
A. Bhaskar, Z.-R. Lin, and C.-J. Liu, 
J. Mater. Sci. {\bf 49}, 1359 (2014).

\bibitem{Limelette2010}
P. Limelette, W. Saulquin, H. Muguerra, and D. Grebille, 
Phys. Rev. B {\bf 81}, 115113 (2010).



\bibitem{Hsieh2014}
Y.-C. Hsieh, R. Okazaki, H. Taniguchi, and I. Terasaki,
J. Phys. Soc. Jpn. {\bf 83}, 054710 (2014).

\bibitem{Sugiyama2002}
J. Sugiyama, H. Itahara, T. Tani, J. H. Brewer, and E. J. Ansaldo, 
Phys. Rev. B {\bf 66}, 134413 (2002).





\bibitem{Koshibae2000}
W. Koshibae, K. Tsutsui, and S. Maekawa,
Phys. Rev. B {\bf 62}, 6869 (2000).

\bibitem{Koshibae2001}
W. Koshibae and S. Maekawa,
Phys. Rev. Lett. {\bf 87}, 236603 (2001).

\bibitem{Sugiyama2003}
J. Sugiyama, J. H. Brewer, E. J. Ansaldo, H. Itahara, K. Dohmae, Y. Seno, C. Xia, and T. Tani,
Phys. Rev. B {\bf 68}, 134423 (2003).

\bibitem{Wakisaka2008}
Y. Wakisaka, S. Hirata, T. Mizokawa, Y. Suzuki, Y. Miyazaki, and T. Kajitani,
Phys. Rev. B {\bf 78}, 235107 (2008).


\bibitem{Raccah1967}
P. M. Raccah and J. B. Goodenough, 
Phys. Rev. {\bf 155}, 932 (1967). 

\bibitem{Asai1989}
K. Asai, P. Gehring, H. Chou, and G. Shirane, 
Phys. Rev. B {\bf 40}, 10982 (1989).

\bibitem{Korotin1996}
M. A. Korotin, S. Y. Ezhov, I. V. Solovyev, V. I. Anisimov, D. I. Khomskii, and G. A. Sawatzky, 
Phys. Rev. B {\bf 54}, 5309 (1996).

\bibitem{Haverkort2006}
M. W. Haverkort, Z. Hu, J. C. Cezar, T. Burnus, H. Hartmann, M. Reuther, C. Zobel, T. Lorenz, A. Tanaka, N. B. Brookes, H. H. Hsieh, H.-J. Lin, C. T. Chen, and L. H. Tjeng, 
Phys. Rev. Lett. {\bf 97}, 176405 (2006).


\bibitem{Klein2006}
Y. Klein, S. H\'ebert, D. Pelloquin, V. Hardy, and A. Maignan,
Phys. Rev. B {\bf 73}, 165121 (2006).

\bibitem{Shibasaki2011}
S. Shibasaki, I. Terasaki, E. Nishibori, H. Sawa, J. Lybeck, H. Yamauchi, and M. Karppinen,
Phys. Rev. B {\bf 83}, 094405 (2011).

\bibitem{Okazaki2011}
R. Okazaki, Y. Nishina, Y. Yasui, S. Shibasaki, and I. Terasaki,
Phys. Rev. B {\bf 84}, 075110 (2011).


\bibitem{Huang2013}
Y. Huang, B. Zhao, R. Ang, S. Lin, Z. Huang, S. Tan, Y. Liu, W. Song, and Y. Sun,
J. Phys. Chem. C {\bf 117}, 11459 (2013).


\bibitem{Mikami2006} 
M. Mikami, K. Chong, Y. Miyazaki, T. Kajitani, T. Inoue, S. Sodeoka, and R. Funahashi, 
Jpn. J. Appl. Phys. {\bf 45}, 4131 (2006).


\bibitem{Kadowaki1986}
K. Kadowaki and S. B. Woods, 
Solid State Commun. {\bf 58}, 507 (􏰀1986􏰁).

\bibitem{Shibasaki2006}
S. Shibasaki and I. Terasaki,
J. Phys. Soc. Jpn. {\bf 75}, 024705 (2006).

\bibitem{Takahashi2012}
R. Takahashi, R. Okazaki, Y. Yasui, I. Terasaki, T. Sudayama, H. Nakao, Y. Yamasaki, J. Okamoto, Y. Murakami, and Y. Kitajima,
J. Appl. Phys. {\bf 112}, 073714 (2012).



\bibitem{Behnia2004}
K. Behnia, D. Jaccard, and J. Flouquet,
J. Phys.: Condens. Matter {\bf 16}, 5187 (2004).


\bibitem{Kresin1990}
V. Z. Kresin and S. A. Wolf,
Phys. Rev. B {\bf 41}, 4278 (1990).

\bibitem{Eng2006}
H. W. Eng, P. Limelette, W. Prellier, C. Simon, and R. Fr\'esard, 
Phys. Rev. B {\bf 73}, 033403 (2006).

\bibitem{Itou2000}
T. Itou and I. Terasaki,
Jpn. J. Appl. Phys.  {\bf 39}, 6658 (2000).


\bibitem{Nakamoto1995}
G. Nakamoto, T. Takabatake, Y. Bando, H. Fujii, K. Izawa, T. Suzuki, T. Fujita, A. Minami, I. Oguro, L. T. Tai, and A. A. Menovsky,
Physica B {\bf 206\&207}, 840 (1995).


\bibitem{Kobayashi2006}
W. Kobayashi and I. Terasaki,
Appl. Phys. Lett. {\bf 89}, 072109 􏰁(2006).􏰀

\bibitem{Wang2010}
Y. Wang, Y. Sui, P. Ren, L. Wang, X. J. Wang, W. H. Su, and H. J. Fan, 
Chem. Mater. {\bf 22}, 1155 (2010).


\bibitem{Furuta2014}
N. Furuta, S. Asai, T. Igarashi, R. Okazaki, Y. Yasui, I. Terasaki, M. Ikeda, T. Fujita, M. Hagiwara, K. Kobayashi, 
R. Kumai, H. Nakao, and Y. Murakami,
Phys. Rev. B {\bf 90}, 144402 (2014).

\bibitem{Takenaka2002}
K. Takenaka, R. Shiozaki, S. Okuyama, J. Nohara, A. Osuka, Y. Takayanagi, and S. Sugai,
Phys. Rev. B {\bf 65}, 092405 (2002).


\bibitem{Pei2012}
Y. Pei, A. D. LaLonde, H. Wang, and G. J. Snyder,
Energy Environ. Sci. {\bf 5}, 7963 (2012) and references therein.

\bibitem{Zhao2014}
L.-D. Zhao, S.-H. Lo, Y. Zhang, H. Sun, G. Tan, C. Uher, C. Wolverton, V. P. Dravid, and M. G. Kanatzidis,
Nature {\bf 508}, 373 (2014).

\end{thebibliography}
\end{document}